\documentclass[12pt]{article}
\newcommand{\dd}{{\rm d}}
\newcommand{\p}{\partial}
\newcommand{\nnn}{\noindent}
\newcommand{\be}{\begin{equation}}
\newcommand{\ee}{\end{equation}}
\newcommand{\bi}{\bibitem}
\newcommand{\pav}{Pav\v si\v c}

\begin{document}

\

\baselineskip 0.8cm

\vspace{1cm}

\begin{center}{\Large \bf Radiation Reaction and the Electromagnetic Energy-Momentum of Moving Relativistic Charged Membranes}

\vspace{1cm}

A.O.Barut$^{1}$ \footnotetext{$^{1}$Department of Physics, University of Colorado, Boulder, CO 80309, USA} and M.Pav\v si\v c$^{2,3}$ \footnotetext{$^{2}$J.Stefan Institute, University of Ljubljana, Jamova 39, 61000 Ljubljana, Slovenia} \footnotetext{$^{3
}$Presently a visiting scientist at The Institute of Applied Mathematics, IMECC, Unicamp, Brazil}
\\
International Centre for Theoretical Physics, Trieste, Italy

\end{center}

\vspace{3cm}

{\bf Abstract}
\\

The charged membrane of Dirac provides a stable electron model with finite self energy. Its total mass $m$ has been previously calculated from the Hamiltonian of the membrane. To complete the picture we evaluate it here on the basis of the energy-momentum
 of its self field (radiation reaction) and obtain the same result showing the consistency of the model. We show explicitly that the old 4/3-problem does not arise. The electron's stability (the vanishing of total ${T^{\mu \nu}}_{, \nu}$) is due to surfac
e tension $\kappa$ of the membrane, but the model is as simple as the point particle, with two parameters $\kappa$ and
$e$; the surface tension parameter $\kappa$ can be expressed in terms of the mass $m$.

\vspace{4mm}

[Published in {\it Physics Letters} B 331 (1994) 45--50)]

\newpage

\section{Introduction}

In the previous papers \cite{1,2} we have formulated a general theory of moving relativistic charged membranes.
A particularly interesting application is to the Dirac shell model of the electron \cite{3} for which we have found, by
using a more straightforward procedure, the same mass as Dirac. In that approach we assumed a
spherically symmetric, radially oscillating membrane, and then calculated the canonical momentum (which contains the minimal coupling with membrane's own electromagnetic potential)
and the corresponding rest mass. In the present paper we first study the self interaction of the
moving membrane and the correction to the observed membrane's mass due to the radiation reaction force.
So we obtain the observed (or renormalized) mass. Then we calculate, starting from the expression
for the stress-energy tensor, the 4-momentum of the electromagnetic field around a moving membrane to which
we add also the 4-momentum due to membrane's stress. We obtain the same expression for membrane's
mass as before and also as in the previous paper. Thus we verify the consistency of various approaches.

Then we restate and formulate for the case of our spherical moving charged membrane an observation exposed
in refs.\cite{4,4a} concerning the 4-momentum (and mass) of the electromagnetic field around a moving
object. We can perform this calculation by integrating the stress-energy tensor over the space-like
hypersurface which is oriented along the world line of the observer with respect to whom the
membrane is moving. The result is the well known factor of 3/4 for the ratio between the energy of
the electromagnetic field and its inertial mass \cite{5}. This result has been much discussed \cite{5} and the
the researchers of the electromagnetic origin of mass came to the conclusion that not all of the electron
mass is of electromagnetic origin. It has been observed \cite{4,4a} that the integration has to
be done rather over the hypersurface (time slice) which is oriented along the moving particle worldine.
Then one obtains the correct expression for the 4-momentum which is consistent with the Lorentz
transformations.\\

\section{Equations of motion for membrane's centre of mass}

In the following we are going to investigate the motion of membrane's centre of mass. We start
from the action of an electrically charged relativistic membrane of any dimension:

\begin{eqnarray}
 I[x^{\mu}(\xi),A_{\mu}] & = & \int {\rm d}^d {\xi} (\kappa \sqrt{|f|} + e^a {\partial}_a X^{\mu} A_{\mu})
 {\delta}^d (x - X(\xi)) d^D x    \nonumber  \\ 
  &   &+ {1 \over {16 \pi}} \int F_{\mu \nu} F^{\mu \nu} \sqrt{|g|} {\rm d}^d x
\label{1}
\end{eqnarray}

\nnn Here $d$ is the world sheet dimension and $D$ the space-time dimension; ${\xi}^a
, \; a = 0,1,2,...,d-1$ worldsheet coordinates (parameters) and $X^{\mu}(\xi),
\mu = 0,1,2,...,D-1$ the embedding functions, $f_{ab} \equiv {\partial}_a X^{\mu}
\partial_b X_{\mu}$ the induced metric, $f \equiv {\rm det} f_{ab}$ and $e^a$ the
electric charge current density on the world sheet. An action which is equivalent to (\ref{1}) has been written down in eq.(2) of our previous publication \cite{1}.

By varying (\ref{1}) with respect to $X^{\mu}$ we obtain the membrane's equations of motion

\begin{equation}
  \kappa {\partial}_a (\sqrt{|f|} {\partial}^a X^{\mu}) + e^a {\partial}_a X^{\nu} {F_{\nu}}^{\mu} = 0
\label{2}
\end{equation}

\nnn and variation of $A_{\mu}$ gives the Maxwell equations

\be
   {F^{\mu \nu}}_{,\nu} = - 4 \pi j^{\mu} \;\;\; , \;\;\;\; j^{\mu}(x) = 
 \int {\rm d}^d \xi \, e^a {\partial}_a X^{\mu} {\delta}^D (x - X(\xi))
\label{3}
\ee

Let us integrate eq.(\ref{2}) over the world sheet element ${\rm d}^d \xi$  which we can write as ${\rm d}^d \xi = {\rm d} \tau {\rm d} \sigma \; , \; \; \; {\rm d} \sigma = {\rm d} {\sigma}_a n^a$ , where ${\rm d} {\sigma}_a$ is an element of a space-lik
e hypersurface and $n^a$ a normal vector to it. We
obtain

\be
 \kappa \oint {\rm d} {\sigma}_a \sqrt{|f|} {\partial}^a X^{\mu} + \int {\rm d} \tau {\rm d} \sigma e^a {\partial}_a X^{\nu} {F_{\nu}}^{\mu}(x) = 0
\label{4}
\ee

\nnn We identify the total kinetic momentum

\be
  P_{\rm m}^{\mu} \equiv \kappa \int {\rm d} {\sigma}_a \, \sqrt{|f|} \, {\p}^a   X^{\mu}
\label{5}
\ee

\nnn with the centre of mass momentum. We can define the velocity ${\dot X}_{\rm C}^{\mu}$ of the centroid world line as

\be
  P_{\rm m}^{\mu} = {{m {\dot X}_{\rm C}^{\mu}} \over {\sqrt{{\dot X}_{\rm C}^2}}}
\label{5a}
\ee

\nnn where$P_{\rm m}^{\mu} P_{{\rm m} \mu} = m^2$ is the membrane's rest mass. By assuming, as usual, that only the space-like hypersurfaces ${\Sigma}_1$ and ${\Sigma}_2$ do contribute to the left integral of eq.(\ref{4}), and then taking
${\Sigma}_1$ and ${\Sigma}_2$ to be infinitesimally close to each other we obtain from (\ref{4})

\be
 {{{\rm d} P_{\rm m}^{\mu}} \over {{\rm d} \tau}} + \int {\rm d} \sigma\, e^a {\p}_a X^{\nu} {F_{\nu}}^{\mu}(x) = 0
\label{5b}
\ee

\nnn This is the equation of motion for membrane's centre of mass. The electromagnetic field ${F_{\nu}}^{\mu}$ in eq.(\ref{5}) can be taken to consist of the fixed external field ${F_{\nu}}^{\mu {\rm (ext)}}$ and the self-field generated by our membrane:

\be
  F_{\mu \nu}  = F_{\mu \nu}^{\rm (ext)} + F_{\mu \nu}^{\rm (self)}
\label{6}
\ee

\nnn The external field can be expanded arround the centroid world line $X_{\rm C}^{\mu}$:

\be
  F_{\mu \nu}^{\rm (ext)} (X) = F_{\mu \nu}^{\rm (ext)} (X_{\rm C}) + {\p}_{\sigma} {F_{\mu \nu}^{\rm (ext)}|}_{X=X_{\rm C}} (X - X_{\rm C})^{\sigma} + . . .
\label{7}
\ee

\nnn The expression $e^a {\p}_a X^{\nu}$ is equal to $e {\dot X}^{\nu}$, where $e \equiv e^a n_a$ (see \cite{1}), and can be written as $e \dot{X}_{\rm C}^{\nu}
+ e {(\dot{X} - {\dot{X}}_{\rm C})}^{\nu}$. By using (\ref{6}-\ref{7}) the equation of motion becomes

\be
   {{{\rm d} P_{\rm m}^{\mu}} \over {{\rm d} \tau}} + q {\dot{X}}_{\rm C}^{\nu} {F_{\nu}}^{\mu {\rm (ext)}} + {\rm higher \, multipoles} + \int {\rm d} \sigma \, e {\dot X}^{\nu} {F_{\nu}}^{\mu {\rm (self)}} = 0
\label{8}
\ee

\nnn where $q = \int e {\rm d} \sigma$ is the total charge of the membrane.
Eq.(\ref{8}) indicates that the membrane's centre of mass moves approximately as a point particle in a fixed external field; since membrane has actually an extension, there are also contributions due to higher multipoles interaction and the self interacti
on.

In the special case, when there is no external field and the accelleration of membrane's centre of mass iz zero, all the internal forces cancel out, so that the contribution of self force is zero. But if we take non-zero accelleration, then, because of re
tardation, the internal forces do not cancel out. It is such a contribution of self interaction

\be
   K_{{\rm (self)}}^{\mu} = \int {\rm d} \sigma \, e(\sigma) {\dot X}_{\nu}(\xi)
  F^{\mu \nu}(x)
\label{9}
\ee

\nnn that we are going to calculate now.
\\

\section{Membrane's radiation reaction and the renormalization of mass}

The electromagnetic potential around our membrane can be expressed by the
solution of eqs.(\ref{3})

\be
   A^{\mu}(x) = \int e^b ({\xi}') \, \delta \left ( (x - X({\xi}'))^2 \right )
   \, {\p}_b X^{\mu} ({\xi}') {\rm d} {\xi}'
\label{9a}
\ee

\nnn satisfying the Lorenz gauge condition ${\p}_{\mu} A^{\mu} = 0$ . From eq.(\ref{4}) we derive the following expression for the electromagnetic field
tensor $F_{\mu \nu} = {\p}_{\mu} A_{\nu} - {\p}_{\nu} A_{\mu}$ :

\begin{eqnarray}
  F^{\mu \nu}(X(\xi)) & = &\int {\rm d} \sigma \, e(\sigma ') ( {1 \over {R^2}}
  {\left ( X(\xi) - X(\xi ') \right ) }^{[\mu} {\ddot X} {(\xi ')}^{\nu ]}
  - {q \over {R^3}} {\left ( X(\xi) - X(\xi ') \right ) }^{[\mu} {\dot X} {(\xi ')}^{\nu ]}       \nonumber   \\
 &   &+ {1 \over {R^3}} {\left ( X(\xi) - X(\xi,) \right )}^{[\mu}
   {\dot X}^{\nu ]} (\xi ') )
\label{10}
\end{eqnarray}

\be
  R \equiv {\left ( X(\xi) - X(\xi') \right ) }^{\sigma} {\dot X}{(\xi ')}_
 {\sigma} \; , \;  \; \; Q \equiv {\left ( X(\xi) - X(\xi') \right ) }^
 {\sigma }{\ddot X} {(\xi ')}_{\sigma} 
\label{10a}
\ee

\nnn In the expression (\ref{10}) $\xi ' \equiv (\tau ',{\sigma '}^i), \; i = 1,2,...,d-1$ refers to the source point, and therefore $\tau '$ is retarded with respect to $\tau$ of the point at which $F^{\mu \nu}$ is calculated. We now use the expansion

\be
  X^{\mu}(\xi ') = X^{\mu}(\tau + u, \sigma ') = X^{\mu}(\tau) + u {\dot X}^{\mu} + {1 \over 2} u^2 {\ddot X}^{\mu} + {1 \over 6} u^3 {\dot {\ddot X}}^{\mu}
\label{11}
\ee

\nnn and insert it into eq.(\ref{9}) for the self-force. Following the procedure
\footnote{In ref.\cite{6} a much simpler procedure is used to calculated the self interaction of a {\it point particle} with finite electric charge. In the case of our {\it extended object} such a simple technique cannot be used since an infinitesimal
element of the membrane has infinitesimal electric charge; consequently, the integration over the membrane as indicated in eq.(\ref{9}) has to be performed in order to obtain a final contribution.}
analogous to one in \cite{4} we arrive, when the membrane is spherical without oscillations and when its centre of mass speed is much less than the speed of light, at the following result:

\be
  F_{\rm self}^r = - {{q^2} \over {2 r}} {\ddot X}^r + F_{\rm rad}^r + {(\rm higher \, derivatives)} \; , \; \; \; r = 1,2,...,D-1
\label{12}
\ee

\nnn where $r$ is membrane's radius. It has been already found elsewhere \cite{5} that the next term, $F_{\rm rad}^r$, in eq.(\ref{12}) is independent of the size of the extended object:

\be
   F_{\rm rad}^r = {2 \over 3} ({\dot {\ddot X}}^{\mu} + {\dot X}^{\mu}
   {\ddot X}^2)
\label{13}
\ee

\nnn The parameter $\tau$ in eqs.(\ref{12}),(\ref{13}) is now the proper time of the centroid world line. Since the membrane has finite spatial extension, there is also an infinite series of higher derivative terms in eq.(\ref{12}).

The total {\it kinetic momentum} of our spherical membrane can be explicitly calculated from eq.(\ref{5}). We use ${{\rm d} \sigma}_a = n_a {\rm d} \sigma
\; , \; {\rm d} \sigma = {\rm d} {\xi}^1 {\rm d} {\xi}^2 \; , \; n_a {\p}^a X^{\mu} = n^2 {\dot X}^{\mu} \; , \; |f| = |{\bar f} /n^2$ , where ${\bar f} \equiv {\rm det} \, f_{ij}$ is the determinant of the induced metric on the spherical surface. Then, a
fter inserting the membrane's constraint
$n^2 {\dot X}^2 = 1$ , we find

\be
  P_{\rm m}^{\mu} = \kappa \int \dd \sigma \sqrt{\bar f} {{{\dot X}^{\mu}}
  \over {\sqrt{{\dot X}^2}}} = \kappa \, 4 \pi r^2 \,{{{\dot X}^{\mu}}
  \over {\sqrt{{\dot X}^2}}} = m \, {{{\dot X}^{\mu}} \over {\sqrt{{\dot X}^2}}}  
\label{14}
\ee

\nnn where in the last step of eq.(\ref{14}) we assumed a non-oscillating membrane, ${\dot r} = 0$ , so that ${\dot X}^{\mu} = {\dot X}_{\rm C}^{\mu}$ is independent of ${\xi}^1$ , ${\xi}^2$ . From eq.(\ref{14}) we read that the membrane's mass is

\be
   m = 4 \pi r^2 \kappa
\label{15}
\ee

We now insert eqs.(\ref{12}) and (\ref{14}) into the equation of motion (\ref{5}) and identify the coefficient in front of the accelleration as the
renormalized or the experimental (observed) mass:

\be
   m_{\rm exp} = 4 \pi r^2 \kappa + {q^2 \over {2 r}}
\label{16}
\ee

The same result (\ref{16}) for the total membrane's mass - apart from the extra contribution due to radial oscillations) - was obtained in the previos paper \cite{2}. In that work we considered - as a particular case - a spherical charged membrane at rest
 and identified its mass with the time like component of the {\it canonical momentum} integrated over the membrane. We also remarked that when world lines or world sheets are oriented into the future, canonical momentum is equal to the momentum obtained f
rom the stress-energy tensor. But in the case of an antiparticle which has its world line or worldsheet oriented into the past, the two kinds of momenta have oposite signs.\\

\section{The 4-momentum of the electromagnetic field around a moving membrane}

In the following we are going to calculate the momentum obtained from the stress-energy tensor belonging to our membrane's action (\ref{1}):

\be
   T^{\mu \nu} = 2 {{\p {\cal L}} \over {\p g^{\mu \nu}}} = T_{\rm m}^{\mu \nu} +
   T_{\rm EM}^{\mu \nu}
\label{17}
\ee

\be
T_{\rm m}^{\mu \nu} = \kappa \int {\dd}^d \xi \sqrt{|f|} \, {\p}_a X^{\mu} 
{\p}^a X^{\nu} \, {\delta}^D (x - X(\xi))
\label{18}
\ee

\be
   T_{\rm EM}^{\mu \nu} = {1 \over {16 \pi}} F_{\rho \sigma} F^{\rho \sigma} 
   \, g^{\mu \nu} - {1 \over {4 \pi}} F^{\rho \mu} F_{\rho}^{\nu}
\label{19}
\ee

\nnn where m refers to the "matter" and EM to the electromagnetic field contribution. The momentum is equal to the integral of $T_{\mu \nu}$ over a space-like hypersurface:

\be
  P^{\mu} = \int T^{\mu \nu} \dd {\Sigma}_{\nu} = P_{\rm m}^{\mu} + P_{\rm EM}^{\mu}
\label{20}
\ee

For the "matter" part we obtain

\be
  P_{\rm m}^{\mu} =  \int T_{\rm m}^{\mu \nu} \dd {\Sigma}_{\nu} =
  \kappa \int \dd {\sigma}^a \sqrt{|f|} \, {\p}_a X^{\mu}
\label{21}
\ee

\nnn where we have used $\dd {\Sigma}_{\nu} = N_{\nu} \dd {\Sigma} \; , \; \;
  \dd \Sigma \, \dd \tau = {\dd}^D x$ , and we have decomposed
${\p}^a X^{\nu} = n^a \p X^{\nu} + {\bar \p}^a X^{\nu}$ , so that
${\bar{\p}}^a X^{\nu} N_{\nu} = 0$ ,  $\p X^{\nu} N_{\nu} = 1$ ,  $\dd \sigma \,
n^a = \dd {\sigma}^a$ ;  $N_{\nu}$ and $n_a$ are normal vector fields
to our hypersurface in space-time $V_D$ and world sheet $V_d$, respectively,
and $N_{\nu}$ has been taken parallel to membrane's velocity 
$\p X^{\nu} \equiv {\dot X}^{\nu}$. Eq.(\ref{21}) is generally valid and is
indeed equal to the expression (\ref{5}) for the total kinetic momentum. 
If we now consider a particular case of a two dimensional spherical membrane 
which is not oscillating radially, then of course we obtain from (\ref{21}) 
the expression (\ref{14}).

The contribution of the electromagnetic field to the momentum is given by

\be
   P_{\rm EM}^{\mu} = \int T_{\rm EM}^{\mu \nu} \dd {\Sigma}_{\nu}
\label{22}
\ee

\nnn From the expression (\ref{10}) we obtain for the electromagnetic field around a generic membrane

\begin{eqnarray}
  F^{r0} & = & {\dot X}^0 \int \dd \sigma ' e(\sigma') {1 \over R^3} {\left ( X(\xi) - X(\xi ') \right ) }^r - {\dot X}^r \int \dd \sigma ' e(\sigma ') {1 \over R^3}
{\left ( X(\xi) - X(\xi ') \right )}^0   \nonumber \\
 & \simeq & \int \dd \sigma ' e(\sigma') {1 \over R^3} {\left ( X(\xi) - X(\xi ') \right ) }^r \; \; \; {\rm for} \; \; \; v << c = 1
\label{23}
\end{eqnarray}

\begin{eqnarray}
  F^{rs} & = & {\dot X}^r \int \dd \sigma ' e(\sigma') {1 \over R^3} {\left ( X(\xi) - X(\xi ') \right ) }^s - {\dot X}^s \int \dd \sigma ' e(\sigma ') {1 \over R^3}
{\left ( X(\xi) - X(\xi ') \right )}^r   \nonumber  \\
& = &{\dot X}^r F^{s0} - {\dot X}^s F^{r0}
\label{24}
\end{eqnarray}

\nnn For a spherical membrane without oscillations we have

\be
  F^{r0} = {\bf E} = {q \over r^2} \, {{\bf r} \over r}
\label{25}
\ee

\nnn  Now we assume as before that

\be
   \dd {\Sigma}_{\nu} = N_{\nu} \dd \Sigma \; \; , \; \; N_{\nu} = {\dot       X}_{\nu}
\label{26}
\ee

\nnn so that the line element of the hypersurface over which we integrate is oriented in the same direction as membrane's centre of mass 4-velocity ${\dot X}_{\nu}$ . Then, from eqs.(\ref{17}-\ref{26}), we have

\be
P^0 = (4 \kappa \pi r^2 + {q^2 \over {2 r}})
\label{27}
\ee

\be
  {\bf P}^r =  (4 \pi r^2 \kappa + {q^2 \over {2 r}}) {\bf v} \; \; , \; \; \; 
   {\bf v} \equiv v^r = {{\dd x^r} \over {\dd x^0}} \; , \; r = 1,2,3
\label{28}
\ee

\nnn where ${\bf v}$ is membrane's 3-velocity and $q = \int e(\sigma) \dd \sigma$ its total electric charge. Our derivation is valid for the speed $v$ much smaller that the speed of light. However, since $P^0$ is equal to membrane's rest mass, we obtain i
mmediately, by considering a Lorentz transformations, that in general

\be
  P^{\mu} = (4 \pi \kappa r^2 + {q^2 \over {2 r}}) {{\dd x^{\mu}} \over {\dd s}}
\label{28a}
\ee
where $\dd s = \dd x^0 (1 - v^2)^{1/2}$ is an element of the proper time.

In eqs.(\ref{27}-\ref{28a}) we have again the same result for membrane's mass as calculated before in other ways.

Finally, let us observe that in the literature about the electromagnetic origin of electron mass \cite{5} we find the calculations of the momentum of the electromagnetic field which are analogous to one in eqs.(\ref{22}-\ref{28}), except for the choice of
 the hypersurface element $\dd {\Sigma}_{\nu}$. Instead of orienting $\dd {\Sigma}_{\nu}$ along the electron's velocity, they just consider it oriented into the "laboratory" proper time direction:

\be
   \dd {\Sigma}_{\nu} = (\dd {\Sigma}_0 , 0, 0, 0) \; \; , \; \; \dd {\Sigma}_0 = \dd x^1 \dd x^2 \dd x^3
\label{29}
\ee

\nnn The result for membrane's electromagnetic field energy and momentum is then
\cite{5}

\be
  P_{\rm EM}^0 = {q^2 \over {2r}}
\label{30}
\ee

\be
  {\bf P}_{\rm EM} = {2 \over 3} {q^2 \over r} {\bf v} = m_{\rm e} {\bf v}
\label{31}
\ee

\nnn Eqs.(\ref{30}-\ref{31}) represent the well known puzzle of the classical electron theory: electron's electromagnetic energy is equal to 3/4 of the mass
contained in the electromagnetic field:

\be
   P_{\rm EM}^0 = {4 \over 3} m_{\rm el}
\label{32}
\ee

\nnn The well known implication, quoted for example by Feynman \cite{5}, was that not all the electron's mass is contained in its electromagnetic field. Several ways have been proposed to resolve the problem, including the so called Poincar\' e stress.
However, the most serious difficulty is that
the expressions (\ref{30},\ref{31}) are in conflict with Lorentz transformations of $P^{\mu}$. But our starting expressions, like (\ref{22}),(\ref{10}), are all fully Lorentz covariant. Therefore, there must be a mistake in the derivation of energy and mo
mentum as given in (\ref{30}),(\ref{31}). 

The above old calculations have not taken advantage of the modern fully relativistically covariant formalism, and it has escaped to those authors the obvious fact, that covariance of equation (\ref{22}) is maintained only if $\dd {\Sigma}_{\nu}$ is chosen
 consistently \cite{4,4a}; and the result of the integral (\ref{22}) depends on choice of $\dd {\Sigma}_{\nu}$, since on membrane's surface the divergence relation
${{T_{\rm EM}}^{\mu \nu}}_{,\nu} = 0$ is not satisfied \cite{4,4a}. In the literature they inconsistently considered the same hypersurface $\dd {\Sigma}_{\nu} = (\dd {\Sigma}_0, 0, 0, 0)$ when calculating electron's electromagnetic {\it rest} energy $P_{\
rm EM}^0$ and when calculating its {\it momentum} ${\bf P}_{\rm EM}$. However, when particle is at rest, the surrounding magnetic field is zero (see eqs.(\ref{24})) and the momentum is zero. If one transforms $P_{\rm EM}^{\mu} = (P_{\rm EM}^0, 0, 0, 0)$ i
nto another Lorentz frame, then also $\dd {\Sigma}_{\nu} = (\dd {\Sigma}_0, 0, 0, 0)$, which is pointing along particle's world line, transforms accordingly. In the new frame, $\dd {\Sigma}_{\nu}$ again points along the (moving) particle world line, which
 is no more the direction of the "laboratory" or the world line of the observer "at rest".\\

\section{Conclusion}

We have started from general equations of a moving charged membrane and written down the equations of motion of membrane's centre of mass. We have calculated the contribution of the radiation reaction force to a spherical membrane's mass. Then we have wri
tten the general expression for the 4-momentum of the electromagnetic field around a moving membrane. In particular, we calculated the 4-momentum of the field around a spherical membrane. By this two different procedures we obtained the same membrane's ma
ss. Moreover, the resulting membrane's mass coincides with one obtained in \cite{2,3} where yet another different approach was used. So we have now additional confirmation of our treatment of relativistic charged membranes and their self-interaction. We f
eel that such a deep understanding of this self-interactive system is interesting, because it might turn out to be a useful step toward a realistic model of electron with finite self-energy, both in its classical and quantiz!
ed version,
including spin \cite{7,1}.

\vspace{1cm}

\begin{center}{\bf Acknowledgements}
\end{center}

M. Pav\v si\v c wishes to thank prof. A. Salam and ICTP (Trieste, Italy) for the hospitality during the first stage of preparation of this work. He is also grateful
to prof. Waldyr A. Rodrigues, jr. and The Institute of Applied Mathematics, IMECC - UNICAMP (Campinas, Brazil), where this paper was completed, for the hospitality and support. The work was supported by the Slovenian Ministry of Science and Technology, and b
y FAPESP (Brazil).

\end{document}